\documentclass[10pt,aps,prd,showpacs,nofootinbib,floats,floatfix,preprintnumbers,groupedaddress,twocolumn]{revtex4}

\usepackage{aas_macros}
\usepackage{cancel}

\usepackage[normalem]{ulem} 
\usepackage{bm}
\usepackage[titletoc]{appendix}
\usepackage{hyperref}
\usepackage{latexsym}
\usepackage{dcolumn}
\usepackage{amsmath,amsfonts,amssymb}
\usepackage{graphicx,epsfig}
\usepackage{amsmath}
\usepackage{fancyhdr}
\usepackage{hyperref}
\usepackage{graphicx,epstopdf}
\usepackage{xcolor}
\usepackage{siunitx}

\def\bea{\begin{eqnarray}}
	\def\eea{\end{eqnarray}}

\begin{document}

\title{GRMHD modelling of accretion flow around Sagittarius A$^*$ constrained by EHT measurements}
\author{Gargi Sen}\email{g.sen@iitg.ac.in}
\author{Debaprasad Maity}\email{debu@iitg.ernet.in}
\author{Santabrata Das}\email{sbdas@iitg.ac.in (Corresponding Author)}
	
\affiliation{Department of Physics, Indian Institute of Technology Guwahati, Guwahati 781039, Assam, India}
\date{\today}
	
\begin{abstract}

We study low angular momentum, advective accretion flows around a Kerr black hole within the framework of general relativistic magnetohydrodynamics (GRMHD) in the steady state. By solving the full set of GRMHD equations, we aim to provide a comprehensive understanding of the behavior of magnetized plasma in the strong gravity regime near a rotating black hole. The accretion solutions are obtained for a set of input parameters, namely energy ($\mathcal{E}$), angular momentum ($\mathcal{L}$), magnetic flux ($\Phi$), and isorotation parameter ($I$). By systematically varying these parameters, we generate a family of global GRMHD accretion solutions that characterize the physical environment around the black hole. Using this approach, we investigate whether the inferred magnetic field strengths reported by the Event Horizon Telescope (EHT) for Sagittarius A$^*$ at various radii can be reproduced. We find that, for a broad range of parameter values, our model successfully recovers the EHT inferred magnetic field strengths with an accuracy of approximately $10\%$, offering a self-consistent framework for interpreting near-horizon accretion physics.

\end{abstract}
	
\maketitle
	
\section{Introduction}

Black holes, a remarkable prediction of Einstein’s theory of general relativity \cite{Einstein-1916a}, are regions of spacetime where gravity is so extreme that nothing—not even light—can escape their grasp. Although they cannot be observed directly, a variety of powerful indirect methods have been developed to investigate their properties. These include the observation of black hole shadows \cite{Cunningham-Bardeen1973, Luminet-1979, Chandrasekhar-1983}, electromagnetic emissions such as X-rays and radio waves \cite{Schreier-1971, Lynden-Bell-Rees1971, Pringle-1981, Pringle-Rees1972}, and the detection of gravitational waves \cite{Einstein1916, Einstein1918}. Supermassive black holes ($M_{\rm BH} > 10^5 M_{\odot}$) are believed to power quasars and reside at the centers of galaxies, while stellar-mass black holes ($M_{\rm BH} \sim 2$–$10^2 M_{\odot}$) are typically found in X-ray binary systems. The radiation observed from these objects arises from the conversion of gravitational potential energy into electromagnetic energy as matter accretes onto the black hole.

One of the most striking observational signatures of a black hole is its shadow: a dark central region encircled by a bright ring, caused by the bending of light in the intense gravitational field. The source of this light is thought to be the accretion flow—a disk of hot, infalling matter surrounding the black hole—which plays a central role in shaping the observed shadow \cite{Smith-1966, Shakura-Sunyaev1973, Peterson-Bradley1997, Fabian-1999, Martin-2000, Frank-etal2002, Luminet-1979}. In supermassive systems, accretion typically proceeds at low rates, producing hot, radiatively inefficient flows, whereas stellar-mass black holes often accrete matter transferred from a companion star. Understanding the physics of accretion is therefore essential for interpreting the black hole observations and for probing the nature of these extreme astrophysical environments.

Magnetic fields are ubiquitous in astrophysical environments and are expected to play a pivotal role in shaping the dynamics and observational signatures of accretion flows around black holes \cite{Pudritz1981, Dadhich-Wiita1982, Wiita1983, Jacquemin-Ide-etal2024, Zamaninasab-etal2014}. Magnetic fields may originate from the companion star, the surrounding interstellar medium, or be generated internally through dynamo processes within the accretion disk \cite{Torkelsson-Brandenburg1994, Brandenburg-etal1995, Khanna-Camenzind1996, Balbus-Hawley1998a, Balbus-Hawley1998b}. In particular, dynamo mechanism converts the kinetic energy of turbulent motions into magnetic energy \cite{Parker1955, Moffatt1978, Krause-Raedler1980, Molchanov-etal1985}, amplifying and organizing magnetic fields within the accretion flow. Early studies examined the evolution of initially uniform magnetic fields in spherically symmetric accretion onto Schwarzschild black holes, revealing how magnetic energy can reach equipartition with kinetic energy and drive synchrotron emission near the event horizon \cite{Bisnovaty-Ruzmaikin1974, Bisnovatyi-Ruzmaikin1976}. Such emission potentially offers a diagnostic tool to distinguish between spinning and non-spinning black holes. Magnetic turbulence, through Maxwell stresses, is now widely believed to be the primary mechanism for angular momentum transport within accretion disks \cite{Shakura-Sunyaev1973}. The coupling between magnetic fields and differential rotation gives rise to magnetorotational instability (MRI), which generates magnetohydrodynamic (MHD) turbulence, enhances effective viscosity, and enables matter to accrete efficiently \cite{Balbus-Hawley1991}. Furthermore, magnetic fields are crucial in launching bipolar outflows and collimated jets, the features commonly observed in both stellar-mass and supermassive black hole systems \cite{Blandford-Payne1982, Balbus-Hawley1991, Miller-etal2006, Takasao-etal2022, Jana-Das2024}. Similarly, in solar winds, pressure gradients overcome gravity to drive a continuous supersonic plasma outflow, which transports magnetic field lines into interplanetary space forming a spiral structure~\cite{Parker1958, Parker1966}. These magnetized outflows carry away angular momentum via magnetic torques~\cite{Weber-Davis1967}, causing faster rotating stars to spin down more rapidly than slower rotators. Magnetic reconnection processes also contribute to disk heating by dissipating magnetic energy \cite{Kawanakaetal2005, Hirose-etal2006, Machida-etal2006, Blaes-etal2007, Kawanaka-etal2008, Ripperda-etal2020, Camilloni-Rezzolla2024, Shen-YuChih2025}, while synchrotron radiation provides a competing cooling mechanism \cite{Shapiro-etal1983, Chakrabarti-Mandal2006, Sarkar-etal2018}.

Numerical simulations consistently show that, in magnetized accretion disks, the toroidal component of the magnetic field typically dominates over the poloidal component due to the differential rotation of the inflowing material \cite{Stone-Norman1994, Hawley-etal1995, Stone-etal1996, Hawley2001, Kato-etal2003, Hirose-etal2004II, McKinney-Gammie2004, Begelman-Pringle2007, Mishra-etal2020}. Motivated by these findings, numerous attempts were made to study self-consistent global MHD accretion solutions around black holes \cite{Akizuki-Fukue2006, Machida-etal2006, Begelman-Pringle2007, Oda-etal2007, Oda-etal2010, Samadi-etal2013, Sarkar-Das2016, Sarkar-Das2018}. These models often assume a predominantly toroidal magnetic field, consistent with the disk's rotational dynamics. 

In recent past, horizon-scale images of supermassive black holes (SMBHs) have opened an unprecedented observational window to examine the innermost regions of two galaxies. The Event Horizon Telescope (EHT) collaboration produced the first polarimetric images of the black hole at the center of the M87 galaxy, marking a major milestone in black hole astrophysics \cite{EHT-2019I, EHT-2019II, EHT-2019III, EHT-2019IV, EHT-2019V, EHT-2019VI}. Soon after, the EHT collaboration successfully captured the image of Sagittarius A$^*$ (Sgr A$^*$) at the centre of the Milky Way galaxy \cite{EHT-2022I, EHT-2022II, EHT-2022III, EHT-2022IV, EHT-2022V, EHT-2022VI, EHT-2024VII, EHT-2024VIII}.  Imaging and modelling analyses reveal that the image of Sgr A$^*$ is dominated by a bright, thick ring with a diameter of $51.8 \pm 2.3 ~\mu \rm {as}$. To interpret the asymmetric ring, the EHT collaboration employed state-of-the-art general relativistic magnetohydrodynamic (GRMHD) simulations coupled with general relativistic ray-tracing techniques. These simulations explored both Standard and Normal Evolution (SANE) and Magnetically Arrested Disk (MAD) scenarios across a range of black hole spins. The observational data, particularly the resolved polarized structure, favored the MAD framework with a spin parameter of $a_{\rm k} \sim 0.94$, suggesting a magnetically dominated accretion environment \cite{EHT-2022V}. This asymmetry in the ring structure is well explained by strong gravitational lensing of synchrotron radiation emitted by hot plasma near the event horizon, offering compelling evidence for the presence of a Kerr black hole with mass $\sim 4 \times 10^6~M_{\odot}$. The polarimetric measurements provided further constraints on the accretion flow properties, including the black hole spin, the mass accretion rate, the electron-to-ion temperature ratio, and the inclination angle of the observer relative to the angular momentum axis of the flow \cite{EHT-2022V, EHT-2024VII, EHT-2024VIII}. Importantly, the magnetic field configuration near the event horizon was characterized using these data, revealing field strengths of $26^{+3}_{-4}$ G at $7.3~r_g$, $67^{+8}_{-9}$ G at $4~r_g$, and $560^{+80}_{-80}$ G in the immediate vicinity of the event horizon. These measurements, along with signatures of flare-related structural variability, evidently infer the critical role of magnetic fields in the dynamics of the near-horizon region. Simulations based on the MAD scenario have successfully reproduced similar polarimetric features \cite{Palumbo-etal2023, Narayan-etal2021}, demonstrating the dynamic role of strong magnetic fields in shaping the observed emissions from SMBHs.

Motivated by these groundbreaking observations, we aim to investigate the accretion dynamics around a Kerr black hole by solving the ideal GRMHD equations \cite{Mitra-Das2024}, constrained by the magnetic field strengths at various radii as reported by the EHT for Sgr A$^*$. Our model framework focuses on key flow parameters, namely energy ($\mathcal{E}$), angular momentum ($\mathcal{L}$), magnetic flux ($\Phi$), and isorotation parameter ($I$) \cite{McKinney-Gammie2004}, to construct self-consistent GRMHD accretion solutions. We incorporate a relativistic equation of state to accurately describe the thermodynamic properties of the inflowing plasma \cite{Chattopadhyay-Ryu2009}. By systematically exploring the $\mathcal{L}–\Phi$ parameter space, we identify accretion solutions that reproduce the EHT inferred magnetic field strengths with an accuracy of approximately $10\%$, providing valuable insights into the structure and magnetization of near-horizon accretion flows.

This paper is organized as follows. In \S II, we describe the GRMHD framework and underlying model assumptions. In \S III, we present the obtained results and discuss the astrophysical implication of our model. Finally, we summarize our findings in \S IV.

\section{GRMHD Framework and Model Assumptions}

We examine the accretion flow around a Kerr black hole in the GRMHD framework. The line element for a stationary, axisymmetric spacetime is expressed as,
\begin{align}\label{eq:line}
    \begin{split}
    ds^2 = & g_{\mu\nu}dx^\mu dx^\nu\\
    =& g_{tt}dt^2 + g_{rr}dr^2 + 2 g_{t\phi} dt d\phi + g_{\phi\phi} d\phi^2 +g_{\theta\theta} d\theta^2.
    \end{split}
\end{align}
In terms of Boyer-Lindquist coordinates \cite{Boyer-etal1967}, the components of the Kerr metric are given by,
\begin{align}\label{eq:metric}
    \begin{split}
    g_{tt} &= (a_{\rm k}^{2}\sin^{2}\theta-\Delta)/\Sigma,\\
    g_{t\phi} &= (A\Delta-a_{\rm k}B\sin^{2}\theta)/\Sigma,\\ 
    g_{rr} &= \Sigma/\Delta, \\ 
    g_{\theta\theta} &= \Sigma,\\ 
    g_{\phi\phi} &= (B^{2}\sin^{2}\theta-A^{2}\Delta)/\Sigma, 
    \end{split}
\end{align} 
where $A= a_{\rm k}\sin^{2}\theta$, $\Sigma = a_{\rm k}^2\cos\theta^2+r^2$, $B= r^2+a^2_{\rm k}$, and $\Delta= (r-r_{\rm H})(r-r_{\rm C})$. Here, $r_{\rm H} ~(=1+\sqrt{1-a_{\rm k}^2})$ is the event horizon, $r_{\rm C} ~(=1-\sqrt{1-a_{\rm k}^2})$ is the Cauchy horizon, and $a_{\rm k}$ is the Kerr parameter. In our analysis, we use ($-, +, +, +$) sign convention and adopt a unit system as $M_{\textrm{BH}} = G = c = 1$, where $M_{\textrm{BH}}$ is the mass of the black hole, $G$ is the universal gravitational constant, and $c$ is the speed of light. In this unit system, length, angular momentum, and energy are expressed in terms of $r_g~(= GM_{\textrm{BH}}/c^{2})$, $r_g c$ and $c^{2}$, respectively. We constrain our whole analysis on the equatorial plane of the disk considering $\theta=\pi/2$.

\subsection{Governing Equations}

We use fundamental principles to obtain the GRMHD equations that describe a relativistic, magnetized, advective accretion flow around a rotating black hole in the steady state. These equations are derived from the conservation laws, $i.e.,$ the conservation of mass, conservation of energy-momentum, and homogeneous Faraday's law in the presence of strong gravitational fields \cite{anile1989,Villiers-2003,Gammie-etal2003,McKinney-etal2004}, which are as follows:
\begin{equation}\label{eq:conserved}
    \nabla_\mu \left(\rho u^\mu \right) = 0;~
    \nabla_\mu T^{\mu\nu}=0;~
    \nabla_\mu {}^*F^{\mu\nu}=0.
\end{equation}
Here, $\rho$ is the mass density and $u^{\mu}$ is the four-velocity of matter. $T^{\mu\nu}$ is the energy-momentum tensor that describes the matter distributions. $^{*}F^{\mu\nu}= \frac{1}{2}(-g)^{-1/2}\eta^{\mu\nu\delta\kappa}F_{\delta\kappa}$ is the Hodge dual of the Faraday electromagnetic tensor $F^{\mu\nu}$ and $\eta^{\mu\nu\delta\kappa}$ is the Levi-Civita symbol. Considering the fluid and Maxwell part, the total energy-momentum tensor is given by,   
\begin{equation}\label{eq:energy-momentum1}
    T^{\mu\nu}=(e+p_{\rm gas})u^\mu u^\nu + p_{\rm gas} g^{\mu\nu} + F^\mu_{~\lambda} F^{\nu \lambda} - \frac{1}{4} F^2 g ^{\mu\nu},
\end{equation}
where $e$ is the local energy density,  $p_{\rm gas}$ is the gas pressure of the flow, and $F^2=F_{\mu \nu}F^{\mu \nu}$. 
We consider perfectly conducting fluid (ideal MHD condition)~\cite{McKinney-etal2004, Baumgarte-Shapiro2002}, where the electric field in the rest frame of fluid is zero, which implies $e^\mu = F^{\mu\nu}u_\nu=0$. In this frame, the magnetic field is written as $b^\mu = {}^*F^{\mu\nu}u_{\nu}$, where $u_{\mu}b^{\mu}=0$. Finally, we write the electromagnetic tensor as
$F^{\mu\nu}= - (-g)^{-1/2}\eta^{\mu\nu\lambda\delta}u_\lambda b_\delta$. Using the aforementioned relations, we get the energy-momentum tensor as,
\begin{equation}\label{eq:energy-momentum2}
    T^{\mu\nu}=
    \rho h_{\rm tot} u^\mu u^\nu + p_{\rm tot} g^{\mu \nu} - b^\mu b^{\nu},
\end{equation}
where $h_{\rm tot}~(= h_{\rm gas} + \frac{B^2}{\rho})$ is the total specific enthalpy of the fluid, and $h_{\rm gas}=(e+p_{\rm gas})/\rho$. The total pressure is $p_{\rm tot} = p_{\rm gas} + p_{\rm mag}$ with $p_{\rm mag}=B^2/2$ and $B^2= b_\mu b^\mu$. We define plasma-$\beta$ as the ratio of gas pressure to magnetic pressure ($\beta = p_{\rm gas}/p_{\rm mag}$) that measures the magnetic activity inside the disk.

\subsection{Conserved Quantities}

Using Eq.~\ref{eq:conserved}, we get the conserved mass accretion rate along the radial direction as,
\begin{equation}\label{eq:mass cons}
    \dot{M} = -4\pi \rho v \gamma_v  H \sqrt{\Delta}=\rm{Constant},
\end{equation}
where ${\dot M}$ refers the mass accretion rate, and $H$ denotes the local half-thickness of the disk. In this work, we consider a $1.5$-dimensional flow model \cite[]{Chakrabarti-1989}, in which the disk is assumed to remain in vertical hydrostatic equilibrium. Following \citep{Riffert-Herold1995}, we compute $H$, which is given by,
$$
    H^2 = \frac{p_{\rm gas}r^3}{\rho \mathcal{F}},~~\mathcal{F}=\gamma_\phi^2\frac{(r^2 + a_k^2)^2 + 2\Delta a_k^2}{(r^2 + a_k^2)^2 - 2\Delta a_k^2}.
$$
Here, $v$ is the radial three-velocity in the co-rotating frame and is defined as $v^{2}=\gamma^{2}_{\phi}v^{2}_{r}$ with $\gamma^{2}_{\phi}=1/(1-v^{2}_{\phi})$, $v^{2}_{\phi}=(u^{\phi}u_{\phi})/(-u^{t}u_{t})$, and $v^{2}_{r}=(u^{r}u_{r})/(-u^{t}u_{t})$. The radial Lorentz factor is $\gamma_{v} = 1/(1-v^{2})^{1/2}$ and $\Omega= u^{\phi}/u^{t}=(g^{t \phi}-\lambda g^{\phi \phi})/(g^{tt}-\lambda g^{t \phi})$ is the angular velocity of the fluid.

The Kerr spacetime possesses two commuting Killing vectors associated with time ($\xi^{t}$) and azimuthal coordinate ($\xi^{\phi}$). Using $\nabla_\mu (T^{\mu \nu} \xi_{\nu})=0$, we derive two conserved quantities along $t$ and $\phi$ directions. Along the radial direction, we get,
\begin{equation}\label{eq:energy}
    \frac{-\sqrt{-g} \hspace{0.3cm}  T^r_t}{\sqrt{-g} \rho u^r}  = - h_{\rm tot} u_t +  \frac{1}{\rho u^r} b^r \big(g_{tt} b^t + g_{t\phi} b^\phi \big)=\mathcal{E},
\end{equation}
and
\begin{equation}\label{eq:angular momen}
   \frac{\sqrt{-g} \hspace{0.3cm}  T^r_\phi}{\sqrt{-g} \rho u^r}= h_{\rm tot} u_\phi -  \frac{1}{\rho u^r} b^r \big(g_{\phi\phi} b^\phi + g_{t\phi} b^t \big)= \mathcal{L},
\end{equation}
where $\mathcal{E}$ and $\mathcal{L}$ are the conserved energy and angular momentum, respectively. The time component of the source-free Maxwell equation (Eq.~\ref{eq:conserved}) implies,
\begin{equation}\label{eq:flux}
    -\sqrt{-g}~{}^*F^{r t} = \sqrt{-g} (u^t b^r - u^r b^t)=\Phi,
\end{equation}
and $\phi$ component of the equation (Eq.~\ref{eq:conserved}) gives us the relativistic isorotation equation \cite{McKinney-etal2004} as,
\begin{equation}\label{eq:isorotation}
    \sqrt{-g} ~{}^*F^{r\phi}=\sqrt{-g} (u^r b^\phi - u^\phi b^r)=I.
\end{equation}
Here, $\Phi$ and $I$ are the magnetic flux and isorotation parameter, respectively, that remain constant all throughout along the flow streamline within the disk.

Next, we project the energy-momentum conservation equation (Eq.~\ref{eq:conserved}) along the three spatial directions using the projection operator $h^i_\mu = \delta^i_\mu + u^i u_\mu$, where $i = 1, 2, 3$ and $\mu, \nu = 0, 1, 2, 3$. This yields the component form of the Navier–Stokes equations. The $r$-component of the Navier-Stokes equation, equivalently the radial momentum equation, is given by,
\begin{equation}\label{eq:radial momen1}
    h^r_\mu \nabla_\nu T^{\mu\nu}=0.
\end{equation}

It is worth mentioning that in the immediate vicinity of the black hole, the accretion flow is inherently relativistic due to its bulk velocity and thermal energy, whereas at sufficiently large radii, the flow is essentially non-relativistic. This implies that as matter accretes inward, it undergoes a natural transition from the non-relativistic to the relativistic regime. Describing such trans-relativistic flows requires a relativistically consistent equation of state (EoS), since formulations based on a fixed adiabatic index $\Gamma$ are inadequate for capturing the thermodynamic behavior across these regimes. Therefore, to close the system of governing equations, we employ the Relativistic Equation of State (EoS), where the adiabatic index ($\Gamma$) depends on the temperature ($T$) and the fluid composition. Following \cite{Chattopadhyay-Ryu2009}, the local energy density ($e$), gas pressure ($p_{\rm gas}$) and mass density ($\rho$) are related as,
\begin{equation}
    e = \frac{\rho f}{\tau}, \quad p_{\rm gas}=\frac{2\rho\Theta}{\tau},
    \label{eos}
\end{equation}
where 
\begin{equation*}
    \begin{aligned}
    f=(2-\zeta)&\left[1+\Theta\left(\frac{9\Theta+3}{3\Theta+2}\right)\right]\\
    &+\zeta\left[\frac{m_p}{m_e}+\Theta\left(\frac{9\Theta m_e+3m_p}{3\Theta m_e+2m_p}\right)\right],
\end{aligned}
\end{equation*}
with $\tau = 1+(m_p/m_e)$, $\zeta=n_p/n_e$, and dimensionless temperature $\Theta~=k_{\rm B}T/m_e c^2$. Here, $m_p$ and $m_e$ denote the masses of the ion and electron, respectively, while $n_p$ and $n_e$ represent their corresponding number densities. In this study, we consider $\zeta=1$ for simplicity. The polytropic index ($N$) and the specific heat ratio ($\Gamma$) are given by $N = \frac{1}{2} \frac{df}{d\Theta}$ and $\Gamma = 1 + \frac{1}{N}$. It is noteworthy that the characteristic wave speeds in magnetized flows correspond to the \text{Alfv\'en} and magnetosonic waves. Accordingly, we define the \text{Alfv\'en} speed as $C_a^2 = B^2 / (\rho h_{\rm tot})$ and the fast magnetosonic speed as $C^2_{\rm f}=C_s^2 + C_a^2 - C_s^2 C_a^2$ \cite[see][]{Gammie-etal2003}, where the relativistic sound speed is expressed as $C_s^2 = \Gamma p_{\rm gas} / \rho h_{\rm gas}$ and the magnetosonic Mach number is given by $M =v/C_{\rm f}$.

Next, using the condition $b^{\mu} u_{\mu} = 0$, we determine the components of the magnetic field as follows:
\begin{align}\label{eq:mag components}
\begin{split}
    b^\phi &= \frac{\frac{I}{\sqrt{-g}} + \frac{\Omega \Phi}{\sqrt{-g} \left( 1 - v_r^2 \right)}}{\frac{v\gamma_v}{\sqrt{g_{rr}}} \left( 1 - \frac{\lambda \Omega}{1 - v_r^2} \right)}, \\ 
    b^r &= \frac{\frac{\Phi}{\sqrt{-g}} + \lambda b^\phi \frac{v\gamma_v}{\sqrt{g_{rr}}}}{\left(\gamma_{\phi}\gamma_v\sqrt{\frac{-1}{g_{tt}+g_{t\phi}\Omega}}\right) \left( 1 - v_r^2 \right)},\\
    b^t &= -b^r \frac{v\sqrt{g_{tt}+g_{t\phi}\Omega}}{\gamma_{\phi}\sqrt{-g_{rr}}} + \lambda b^\phi.
\end{split}
\end{align}
With this, the total strength of the magnetic field is obtained as,
\begin{equation}\label{eq:tot mag}
    B=\Bigl(g_{rr}(b^r)^2+g_{tt}(b^t)^2+g_{\phi\phi}(b^{\phi})^2+2
    g_{t \phi}b^tb^{\phi}\Bigr)^{\frac{1}{2}}.
\end{equation}

\subsection{Critical Points Analysis and Transonic Solutions}

The derivatives of the energy conservation equation (Eq.~\ref{eq:energy}) and angular momentum conservation equation (Eq.~\ref{eq:angular momen}) yield,
\begin{equation}\label{eq:energy con}
    \frac{d\mathcal{E}}{dr}=\mathcal{E}_r+\mathcal{E}_v \frac{d v}{dr}+\mathcal{E}_{\Theta}\frac{d \Theta}{dr}+\mathcal{E}_{\lambda}\frac{d \lambda}{dr}=0,
\end{equation}
\begin{equation}\label{eq:angular momen con}
    \frac{d\mathcal{L}}{dr}=\mathcal{L}_r+\mathcal{L}_v \frac{d v}{dr}+\mathcal{L}_{\Theta}\frac{d \Theta}{dr}+\mathcal{L}_{\lambda}\frac{d \lambda}{dr}=0.
\end{equation}
We rewrite the radial momentum equation (Eq.~\ref{eq:radial momen1}) as,
\begin{equation}\label{eq:radial momen2}
    \mathcal{R}_r+\mathcal{R}_v \frac{d v}{dr}+\mathcal{R}_{\Theta}\frac{d \Theta}{dr}+\mathcal{R}_{\lambda}\frac{d \lambda}{dr}=0.
\end{equation}
All coefficients in Eqs.~\ref{eq:energy con}–\ref{eq:radial momen2}, namely $\mathcal{E}_j$, $\mathcal{L}_j$, and $\mathcal{R}_j$ (with $j = r, v, \Theta, \lambda$), are provided explicitly in the Appendix \ref{appen}.
Combining Eqs.~\ref{eq:energy con}-\ref{eq:radial momen2}, we obtain the wind equation of the GRMHD flow as,
\begin{equation}\label{eq:vel grad}
    \frac{dv}{dr}=\frac{-(\mathcal{R}_r +\mathcal{R}_\Theta \Theta_r + \mathcal{R}_\lambda \lambda_r )}{(\mathcal{R}_v +\mathcal{R}_\Theta \Theta_v + \mathcal{R}_\lambda \lambda_v )} = \frac{\mathcal{N}(r,v,\Theta,\lambda)}{\mathcal{D}(r,v,\Theta,\lambda)},
\end{equation}
where 
\begin{align*}
    \Theta_r=&(\mathcal{E}_\lambda \mathcal{L}_{r}-\mathcal{E}_r \mathcal{L}_{\lambda})/(\mathcal{E}_\Theta \mathcal{L}_{\lambda}-\mathcal{E}_{\lambda} \mathcal{L}_{\Theta}), \\
    \Theta_{v}=&(\mathcal{E}_\lambda \mathcal{L}_{v}-\mathcal{E}_v \mathcal{L}_{\lambda})/(\mathcal{E}_\Theta \mathcal{L}_{\lambda}-\mathcal{E}_{\lambda} \mathcal{L}_{\Theta}),\\
    \lambda_{r}=&(\mathcal{E}_\Theta \mathcal{L}_{r}-\mathcal{E}_r \mathcal{L}_{\Theta})/(\mathcal{E}_{\lambda} \mathcal{L}_{\Theta}-\mathcal{E}_\Theta \mathcal{L}_{\lambda}),\\
    \lambda_v=&(\mathcal{E}_\Theta \mathcal{L}_{v}-\mathcal{E}_v \mathcal{L}_{\Theta})/(\mathcal{E}_{\lambda} \mathcal{L}_{\Theta}-\mathcal{E}_\Theta \mathcal{L}_{\lambda}),
\end{align*}
and $\mathcal{N}(r,v,\Theta,\lambda)$ is the numerator and $\mathcal{D}(r,v,\Theta,\lambda)$ is the denominator, respectively. Furthermore, the gradient temperature and angular momentum are expressed in terms of $\frac{dv}{dr}$ as,
\begin{equation}\label{eq:temp grad}
    \begin{split}
    \frac{d\Theta}{dr}= \Theta_r + \Theta_v\frac{dv}{dr},
    \end{split}
\end{equation}
and
\begin{equation}\label{eq:lm grad}
    \begin{split}
    \frac{d\lambda}{dr}=\lambda_r + \lambda_v\frac{dv}{dr},
    \end{split}
\end{equation}
In the presence of magnetic fields, we simultaneously solve Eqs.~\ref{eq:vel grad}–\ref{eq:lm grad} for a set of model parameters ($\mathcal{E}, \mathcal{L}, \Phi, I$) to obtain the GRMHD accretion flow solutions around a Kerr black hole of spin $a_{\rm k}$.

The black hole accretion process initiates from the distant outskirts of the accretion disk, where gravitational influence begins to dominate the dynamics of the inflowing matter. At the outer edge of the accretion disk ($r_{\rm edge}$), the accreting plasma exhibits negligible radial motion  ($v << 1$) and remains subsonic, setting the stage for a transonic transition as it spirals inward toward the event horizon. As the flow spirals inward, the immense gravitational pull of the black hole accelerates the material, causing the radial velocity to rise rapidly. Near the event horizon ($r_{\rm H}$), the inflow velocity asymptotically approaches the speed of light ($v \sim 1$), satisfying the inner boundary conditions. The gravitational attraction of the black hole causes the inflowing matter to undergo a transonic transition, changing from subsonic to supersonic flow. This transition necessitates that the flow passes smoothly through a critical point ($r_c$) on its way to the event horizon. At the critical point, the denominator ($\mathcal{D}$) of Eq.~\ref{eq:vel grad} becomes zero. Since the accretion flow remains smooth and continuous outside the event horizon, the numerator ($\mathcal{N}$) must also vanish at $r_c$ to maintain regularity. As a result, Eq.~\ref{eq:vel grad} takes the indeterminate form $\frac{dv}{dr}\big|_{r_c} = \frac{\mathcal{N}}{\mathcal{D}} \big|_{r_c} = \frac{0}{0}$. To resolve this, we apply l$'$H\^{o}pital's rule to evaluate the velocity gradient at the critical point at $r_c$. In this study, we focus on accretion solution that possesses critical point where $\frac{dv}{dr}|_{r_c}$ owns two real values with opposite signs \cite[]{kato-etal1993, Chakrabarti-Das2004}. It is worth mentioning that such solutions have been demonstrated to be stable under perturbations \cite{kato-etal1993}. Among the two real values, the negative value of $\frac{dv}{dr}|_{r_c}$ corresponds to the accretion solution, while the positive value corresponds to the winds \cite[]{Das-2007,Sarkar-Das2016}. In the present work, we focus exclusively on accretion solutions.

Depending on the input parameters, the accretion flow may contain single or multiple critical points. A critical point located near the horizon is termed the inner critical point ($r_{\rm in}$), while one formed far away from the horizon is called the outer critical point ($r_{\rm out}$) \cite[and references therein]{Chakrabarti-Das2004,Das-2007}. By choosing the set of input parameters ($\mathcal{E}$, $\mathcal{L}$, $\Phi$, $I$, $a_{\rm k}$, and $\dot{M}$), we simultaneously solve $\mathcal{N}=0$ and $\mathcal{D}=0$ along with Eq.~\ref{eq:energy} and Eq.~\ref{eq:angular momen} to compute the flow variables, such as radial velocity ($v_{\rm c}$), temperature ($\Theta_{\rm c}$), and angular momentum ($\lambda_{\rm c}$) at $r_{\rm c}$. Using these flow variable, we integrate Eqs.~\ref{eq:vel grad}–\ref{eq:lm grad} outward from $r_{\rm c}$ to the outer edge ($r_{\rm edge}$) and inward to the black hole horizon ($r_{\rm H}$). Combining these two branches yields a global accretion solution for the GRMHD flow around rotating black hole \cite{Mitra-etal2022,Mitra-Das2024}.

The formalism developed in this study to obtain global GRMHD accretion solutions serves as a robust framework to investigate the magnetic environment of Sgr A$^*$. As the supermassive black hole at the center of our galaxy, Sgr A$^*$ exhibits low-luminosity accretion, where magnetic fields are expected to play a pivotal role in shaping the structure and dynamics of the accretion flow. By adopting suitable input parameters consistent with observational constraints for Sgr A$^*$, the present model framework allows for a self-consistent analysis of the flow structure, including the profiles of Mach number ($M$), velocity ($v$), density ($\rho$), temperature ($T$), magnetic field ($B$) and plasma-$\beta$. This, in turn, can provide valuable insights into the physical conditions prevailing in the vicinity of the event horizon and contribute to the theoretical interpretation of high-resolution observations, such as those conducted by the Event Horizon Telescope (EHT) collaboration~\cite{EHT-2024VII, EHT-2024VIII}.

\section{Results and Astrophysical Implications}

In this section, we explore the astrophysical relevance of our model solutions, with a particular focus on the compact, luminous source Sgr A$^*$. This supermassive black hole, located at the center of the Milky Way, is a prominent emitter of radio waves \cite{Balick-Brown1974, Ekers-etal1975}. Owing to its distinctive characteristics, Sgr A$^*$ serves as a natural laboratory for testing the predictions of general relativity. Long-term observational campaigns, encompassing precise measurements of its proper motion and the orbital dynamics of nearby stars, have confirmed Sgr A$^*$ as a highly compact mass concentration situated approximately $D \sim 8$ kpc from Earth \cite{Do-etal2019, Gravity-2019}. Furthermore, direct imaging of the central source and its immediate environment by EHT, a global array of eight radio telescopes spanning six geographic locations, has produced an image consistent with the shadow of an accreting Kerr black hole with a mass of $M_{\textrm{BH}} \sim 4 \times 10^6 M_\odot$ \cite{EHT-2022I}. Complementary observations by the Chandra X-ray Observatory, which detect bremsstrahlung emission near the gas capture radius, estimate the accretion rate at large radii ($\sim 10^5 R_S$, where $R_S$ is the Schwarzschild radius) to be in the range of $10^{-6} - 10^{-5} M_{\odot} \mathrm{yr}^{-1}$ \cite{Quataert2002}. In contrast, Faraday rotation measurements of polarized millimeter-wavelength emission near the event horizon suggest a much lower accretion rate, approximately $10^{-9} - 10^{-7} M_{\odot} \mathrm{yr}^{-1}$ \cite{Bower-etal2003, Marrone-etal2005}.

The EHT collaboration investigated MAD models with spin parameters $a_{\rm k} = -0.94, -0.5, 0, 0.5$, and $0.94$, and constrained the mass accretion rate to lie within the range $10^{-9} - 10^{-8} M_{\odot}\rm{yr}^{-1}$, accompanied by an outflow power of approximately $10^{38}~\rm{erg}~\rm{s}^{-1}$. Sgr A$^*$ exhibits broadband emission extending from radio to hard X-ray wavelengths. Comparative analysis of EHT observations and numerical simulations suggests that the mass accretion rate is of the order of $\sim 10^{-8}M_{\odot}\rm{yr}^{-1}$, with a bolometric luminosity not exceeding $10^{36} ~\rm{erg}~\rm{s}^{-1}$ \cite{EHT-2022V}. These findings collectively indicate that Sgr A$^*$ hosts a radiatively inefficient accretion flow, classifying it as a low-luminosity supermassive black hole. Remarkably, EHT imaging reveals a bright, thick emission ring with an angular diameter of $51.8 \pm 2.3~\mu$as \cite{EHT-2022I}. Furthermore, leveraging polarized intensity maps and GRMHD simulations, the EHT collaboration has constrained the radial profile of the magnetic field strength in the vicinity of the event horizon \cite{EHT-2024VII, EHT-2024VIII}. Specifically, the mass-weighted average magnetic field strength is found to be $26^{+3}_{-4}$ G at $7.3r_g$, increasing to $67^{+8}_{-9}$ G at $4r_g$, and reaching $560^{+80}_{-80}$ G near the event horizon. These results are most consistent with MAD models featuring a high spin value of $a_{\rm k} \sim 0.94$, and serve as essential observational benchmarks for validating theoretical models of black holes and their surrounding magnetized accretion flows.

\begin{figure}
    \begin{center}
    \includegraphics[width=\columnwidth]{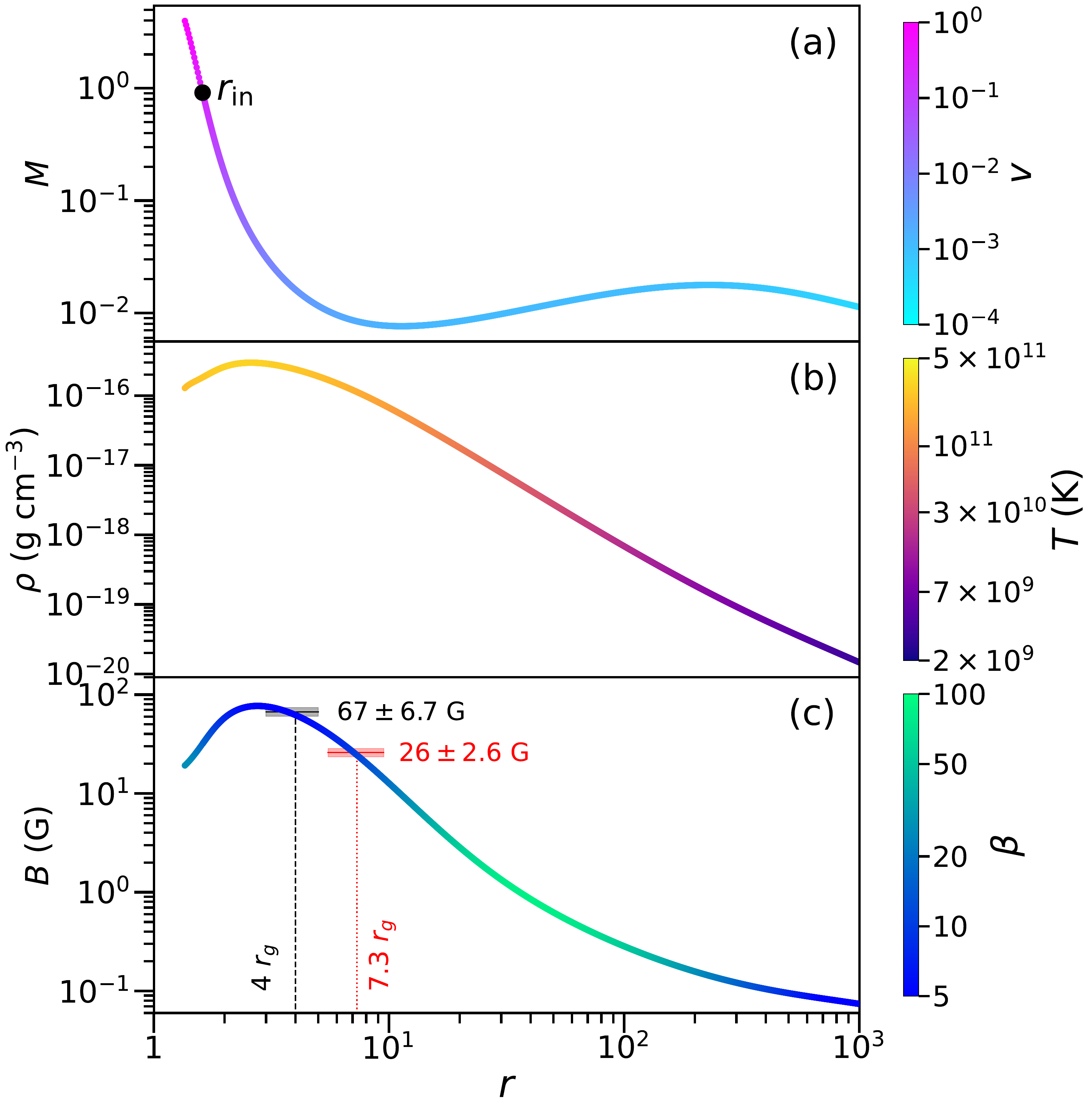}
    \end{center}
    \caption{Example of a global GRMHD accretion solution around a Kerr black hole that accounts the EHT inferred magnetic fields at $7.3~r_{\rm g}$ and $4~r_{\rm g}$. In panel (a), variation of Mach number ($M$) is depicted with radial coordinate ($r$), where color denotes the flow velocity ($v$). Filled circle in black denotes the inner critical point ($r_{\rm in}$). Panel (b) shows the radial variation of density ($\rho$), whereas temperature ($T$) is shown using color. Panel (c) illustrates the magnetic field ($B$) variation with $r$, while plasma-$\beta$ is plotted using color. Thick grey and red horizontal lines denote the magnetic field strengths of $67 \pm 6.7$ G and $26 \pm 2.6$ G, respectively that intersect GRMHD accretion solution at $4~r_{\rm g}$ and $7.3~r_{\rm g}$. See the text for details.
    }
    \label{fig:fig01}
\end{figure}

We therefore aim to investigate the viability of our magnetized accretion solutions, discussed earlier, in light of the recent observational constraints reported by the EHT collaboration~\cite{EHT-2024VIII}. For this purpose, we adopt two key inputs from their analysis, namely the mass accretion rate and the radial profile of the magnetic field strength. In addition, we consider a black hole mass of $M_{\textrm{BH}} = 4 \times 10^6 M_\odot$ \cite{Do-etal2019,Gravity-2019} and a spin parameter $a_{\rm k} = 0.94$ \cite{EHT-2024VIII}, consistent with values reported in the literature. Guided by these, we explore the properties of the corresponding global GRMHD accretion solutions. We depict a representative solution along with its hydrodynamic and magnetic properties in Fig. \ref{fig:fig01}. This solution is selected such that the magnetic field strength remains within $\pm 10 \%$ of the EHT reported best-fit values at $7.3 r_{\rm g}$ and $4 r_{\rm g}$. Here, we choose the following model input parameters as $\mathcal{E}=1.001$, $\mathcal{L}=2.4$, $\Phi=10.5\times10^{-13}$, $I=5\times10^{-15}$, and a mass accretion rate of $\dot{M} = 10^{-8}M_{\odot}\rm{yr}^{-1}$.

In Fig.~\ref{fig:fig01}a, we display the radial variation of the Mach number ($M$), where the color bar indicates the corresponding fluid velocity ($v$). The transition from subsonic to supersonic flow, via the inner critical point located at $r_{\rm in} = 1.612 r_g$ (marked with filled black circle), is clearly evident. The flow velocity at $r_{\rm in}$ is $v(r_{\rm in}) = 0.235$ and the corresponding fast magnetosonic speed at $r_{\rm in}$ is $C_{\rm f}(r_{\rm in}) = 0.255$. Fig.~\ref{fig:fig01}b illustrates the variation of density ($\rho$) as a function of radial distance ($r$), with the color bar representing the temperature of fluid ($T$). As the accreting matter approaches the event horizon, both the density and temperature increase significantly relative to their values at the outer edge ($r_{\rm edge}=10^3 r_g$). The temperature of the accreting plasma exceeds $10^{11}$~K in the innermost regions of the disk, indicative of a geometrically thick, optically thin, and radiatively inefficient hot accretion flow~\cite{Narayan-Yi1994, Yuan-etal2003}. Such extreme thermal conditions are characteristic of low-luminosity accretion systems and are consistent with the physical properties expected near the event horizon of Sgr A$^*$~\cite{Genzel-etal2010, EHT-2022I}. The lower panel (Fig.~\ref{fig:fig01}c) presents the variation of the total magnetic field strength ($B$) as the accretion flow progresses inward toward the horizon. The associated color bar represents the plasma-$\beta$ parameter, which quantifies the ratio of gas pressure to magnetic pressure, thereby offering insights into the relative dominance of magnetic pressure over thermal pressure in different regions of the flow \cite{Mitra-etal2022}. Overplotted on this panel are two horizontal shaded bands in red and grey that correspond to the mean magnetic field strengths of $26$~G and $67$~G, respectively, each with $10\%$ uncertainty, as inferred from the EHT observations. The intersections of these shaded regions with our model GRMHD accretion solution occur at radial distances of approximately $7.3 r_{\rm g}$ and $4 r_{\rm g}$ that demonstrate the consistency of our model formalism with the findings from EHT collaboration~\cite{EHT-2024VIII} reported for Sgr A$^*$. Note that as matter approaches the black hole horizon, its velocity rises sharply (see panel (a)) due to strong gravity, leading to decreases in density and temperature in order to conserve the mass accretion rate (see Eq. \ref{eq:mass cons})). In an ideal MHD flow with frozen-in magnetic fields, this drop in density with decreasing radius results in a corresponding reduction in magnetic field strength. Furthermore, we assess the state of the magnetized accretion flow by computing the magnetic flux threading the inner region near the black hole using $\int_0^{2\pi}\sqrt{-g}(u^tb^r-u^rb^t) d\phi \sim 1.36\times 10^{25}\,{\rm G}\,{\rm cm^2}$. This flux remains well below the threshold for a magnetically arrested disk (MAD) estimated as $\sim 3.8\times 10^{27}\,{\rm G}\,{\rm cm^2}$ for Sgr A$^*$ \cite{Yuan-Narayan2014}.

\begin{figure}
    \begin{center}
    \includegraphics[width=\columnwidth]{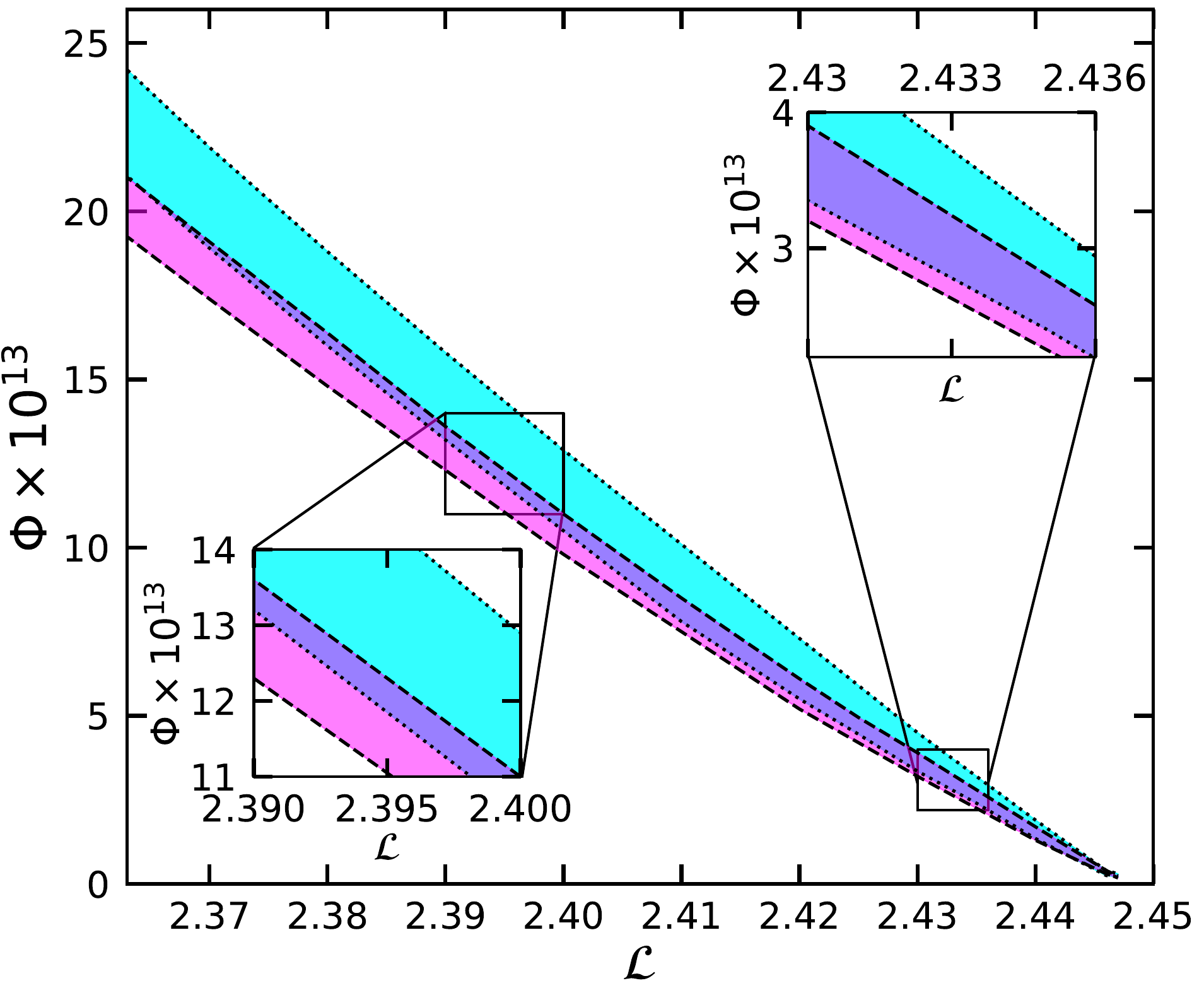}
    \end{center}
    \caption{Effective domain of the parameter space in $\mathcal{L}-\Phi$ plane corresponding to GRMHD accretion solutions that reproduce EHT inferred magnetic field strength. The region shaded with cyan admits accretion solutions yielding magnetic field strength of $67 \pm 6.7$~G at $7.3 r_g$, while magenta shaded region corresponds to solutions matching $26 \pm 2.6$~G at $4r_{\rm g}$. Overlapping region (appeared as violet) provides accretion solutions that simultaneously satisfy magnetic fields constraints at both radii. See the text for details.
    }
    \label{fig:fig02}
\end{figure}

We emphasize that Fig.~\ref{fig:fig01} presents only one representative example among many possible global GRMHD accretion solutions that are consistent with the magnetic field strengths observed by the EHT near Sgr A$^*$ at specific radii. To systematically identify the full range of such GRMHD accretion solutions, we explore the parameter space in the $\mathcal{L} – \Phi$ plane, which reveals a substantial domain supporting magnetic field strengths that match the EHT-inferred values within observational uncertainties. In doing so, we fix the black hole mass $M_{\textrm BH} = 4 \times 10^6M_\odot$ and the mass accretion rate $\dot{M} = 10^{-8}M_{\odot}\rm{yr}^{-1}$, as indicated by EHT observations. Furthermore, we choose $\mathcal{E} = 1.001$, $I = 5\times10^{-15}$ and $a_{\rm k} = 0.94$, as the model parameters. The resulting parameter space is illustrated in Fig.~\ref{fig:fig02}, where we identify effective domains in the $\mathcal{L}–\Phi$ plane that result in magnetic field strengths within $\pm 10\%$ error of the EHT-reported mean values at $4r_{\rm g}$ and $7.3r_{\rm g}$ \cite{EHT-2024VIII}. In the figure, the region shaded in cyan corresponds to GRMHD accretion solutions that yield $B = 67 \pm 6.7$G at $4r_{\rm g}$, while the magenta shaded region includes those solutions consistent with $B = 26 \pm 2.6$G at $7.3r_{\rm g}$. It is important to emphasize that the overlapping region (appeared as violet) in Fig.~\ref{fig:fig02} represents the subset of GRMHD accretion solutions that simultaneously satisfy both magnetic field constraints inferred from the EHT observations, namely, the magnetic field strengths of $B = 26 \pm 2.6$~G at $7.3r_{\rm g}$ and $B = 67 \pm 6.7$~G at $4r_{\rm g}$. The inset panels provide the magnified views of these overlapping regions. This convergence of observational consistency at two distinct radii makes these solutions particularly significant, as they offer a self-consistent framework for describing the magnetized accretion environment around Sgr A$^*$. Therefore, the GRMHD solutions corresponding to this overlapping parameter space are well-suited to interpret and reproduce the magnetic field structure reported by the EHT collaboration~\cite{EHT-2024VIII} for the Galactic Center supermassive black hole Sgr A$^*$. 

\section{Summary and Conclusions}

In this study, we investigate the properties of the magnetized accretion flows around a spinning black hole by solving the GRMHD equations within a simplified 1.5-dimensional disk geometry under steady state conditions. Our motivation stems from recent observations of Sgr A$^*$ by EHT~\cite{EHT-2022I, EHT-2022II, EHT-2022III, EHT-2022IV,EHT-2022V, EHT-2022VI, EHT-2024VII, EHT-2024VIII}, which offer crucial constraints on the magnetic field distribution in the immediate vicinity of the black hole. According to the EHT collaboration, the magnetic field strength near Sgr A$^*$ is estimated to be $26^{+3}_{-4}$G at $7.3r_{\rm g}$ and $67^{+8}_{-9}$G at $4r_{\rm g}$, assuming a Kerr black hole with mass $M_{\textrm{BH}} = 4 \times 10^6 M_\odot$, mass accretion rate $\dot{M} = 10^{-8} M_{\odot}~\rm{yr}^{-1}$, and spin parameter $a_{\rm k} = 0.94$. 

Upon imposing these key constraints, we obtain global GRMHD accretion solutions that are consistent with the reported field strengths within $\pm 10\%$ error at $7.3 r_{\rm g}$ and $4 r_{\rm g}$. One such representative solution is presented that captures the detailed characteristics of the GRMHD accretion flow, including the radial profiles of Mach number ($M$), radial velocity ($v$), density ($\rho$), temperature ($T$), magnetic field strength ($B$), and the plasma-$\beta$ parameter (see Fig.~\ref{fig:fig01}).

To assess the generality of our findings, we systematically explore the parameter space spanned by the magnetic flux ($\Phi$) and the angular momentum ($\mathcal{L}$), identifying the global accretion solutions that remain consistent with the EHT-inferred magnetic field strengths at both radii. This analysis reveals that a family of solutions can simultaneously account for the magnetic field values at $7.3r_{\rm g}$ and $4r_{\rm g}$ as reported by the EHT, suggesting that the magnetized accretion flow around Sgr A$^*$ can be well described by GRMHD models within a constrained but physically realistic range of parameters (see Fig.~\ref{fig:fig02}). With this, we indicate that the present study provides a self-consistent theoretical framework based on GRMHD accretion flows that supports and complements the EHT findings in explaining the magnetized accretion flows around Sgr A$^*$.

At the end, we mention that the present study involves several simplifying assumptions. The accretion flow is modelled in a $1.5$-dimensional geometry, confined near the disk equatorial plane, and the polar magnetic field component ($b^\theta$) is neglected, although it is important for launching winds, jets, and outflows. Radiative cooling is ignored, and the flow is treated as a single-temperature fluid, neglecting the two-temperature nature of ions and electrons. The analysis is also restricted to the ideal MHD limit, omitting resistive or dissipative effects. Implementation of these processes is beyond the scope of the present work and will be addressed in future studies.

\section*{Acknowledgments}

Authors acknowledge the support from the Department of Physics, IIT Guwahati, for providing the facilities to complete this work. Authors also thank Samik Mitra for helpful and insightful discussions during the initial stage of the study.

\appendix

\begin{widetext}

\section{Conservation equations of energy and angular momentum, and radial momentum equation}
\label{appen}

Using the energy conservation equation (Eq.~\ref{eq:energy}), we get, 
\begin{equation*}
    \frac{d\mathcal{E}}{dr}=\mathcal{E}_r+\mathcal{E}_v \frac{d v}{dr}+\mathcal{E}_{\Theta}\frac{d \Theta}{dr}+\mathcal{E}_{\lambda}\frac{d \lambda}{dr}=0.
\end{equation*}

The coefficients $\mathcal{E}_r$, $\mathcal{E}_v$, $\mathcal{E}_\theta$, and $\mathcal{E}_\lambda$ are given by,
\begin{align*}
    \mathcal{E}_r=&- \frac{\partial h_{\rm tot}}{\partial r} u_t -h_{\rm tot} \frac{\partial u_t}{\partial r} +\frac{g_{t\phi}b^r\frac{\partial b^\phi}{\partial r}}{u^r\rho} + \frac{g_{tt}b^r\frac{\partial b^t}{\partial r}}{u^r \rho} \frac{b^r b^t \frac{d g_{tt}}{dr}}{u^r \rho} + \frac{b^r b^{\phi} \frac{d g_{t\phi}}{dr}}{u^r \rho} + \frac{3b^r(g_{tt}b^t+g_{t\phi}b^{\phi})}{2 r u^r \rho}\\
   &+ \frac{\frac{\partial b^r}{\partial r}(g_{tt}b^t+g_{t \phi}b^{\phi})}{u^r \rho}-\frac{\frac{\partial \mathcal{F}}{\partial r} b^r(g_{tt}b^t + g_{\phi\phi} b^{\phi})}{2 \mathcal{F} u^r \rho}  +\frac{b^r\frac{d \Delta}{dr}(g_{tt}b^t+g_{t\phi}b^{\phi})}{2\Delta u^r\rho}-\frac{b^r \frac{\partial u^r}{\partial r}(g_{tt}b^t+ g_{t \phi} b^{\phi})}{(u^r)^2\rho},\\
   \mathcal{E}_v=&-\frac{\partial h_{\rm tot}}{\partial v}u_t -h_{\rm tot}\frac{\partial u_t}{\partial v} + \frac{g_{tt}b^r\frac{\partial b^t}{\partial v}}{u^r \rho}+ \frac{g_{t\phi}b^r\frac{\partial b^{\phi}}{\partial v}}{u^r \rho}+\frac{\frac{\partial b^r}{\partial v}(g_{tt}b^t + g_{t\phi}b^{\phi})}{u^r \rho} -\frac{b^r \frac{\partial u^r}{\partial v}(g_{tt}b^t+g_{t\phi}b^{\phi})}{(u^r)^2 \rho}\\
    &+\frac{b^r(g_{tt}b^t+g_{t\phi})}{u^r \rho v} + \frac{b^r v \gamma_{v}^2 (g_{tt}b^t+g_{t\phi}b^{\phi})}{u^r \rho},\\
    \mathcal{E}_{\Theta}=&-\frac{\partial h_{\rm tot}}{\partial \Theta}u_t + \frac{b^r(g_{tt}b^t+g_{t\phi}b^{\phi})}{r u^r \Theta \rho},\\
    \mathcal{E}_{\lambda}=&-\frac{\partial h_{\rm tot}}{\partial \lambda}u_t -h_{\rm tot}\frac{\partial u_t }{\partial \lambda} + \frac{g_{tt}b^r \frac{\partial b^t}{\partial \lambda}}{u^r \rho} + \frac{g_{t\phi}b^r\frac{\partial b^{\phi}}{\partial \lambda}}{u^r \rho}-\frac{\frac{\partial F}{\partial \lambda}b^r(g_{tt}b^t+g_{t\phi}b^{\phi})}{2 \mathcal{F}u^r \rho}-\frac{\frac{\partial b^r}{\partial \lambda}(g_{tt}b^t+g_{t\phi}b^{\phi})}{u^r \rho}.
\end{align*}

Applying the angular momentum conservation equation (Eq.~\ref{eq:angular momen}), we obtain,
\begin{equation*}
\frac{d\mathcal{L}}{dr}=\mathcal{L}_r+\mathcal{L}_v \frac{d v}{dr}+\mathcal{L}_{\Theta}\frac{d \Theta}{dr}+\mathcal{L}_{\lambda}\frac{d \lambda}{dr}=0,
\end{equation*}
where the coefficients $\mathcal{L}_r, \mathcal{L}_v$, $\mathcal{L}_\theta$ and $\mathcal{L}_\lambda$ are expressed as, 
\begin{align*}
      \mathcal{L}_r =& \frac{\partial h_{\rm tot}}{\partial r}u_{\phi} + h_{\rm tot} \frac{\partial u_{\phi}}{\partial r} -\frac{g_{t \phi}b^r \frac{\partial b^t}{\partial r}}{u^r \rho}-\frac{3 b^r(g_{t\phi}b^t + g_{\phi\phi}b^{\phi})}{2 r u^r \rho} +\frac{\frac{\partial \mathcal{F}}{\partial r}(g_{t\phi}b^t + g_{\phi\phi}b^{\phi})}{2 \mathcal{F}u^r \rho} -\frac{\frac{\partial b^r}{\partial r}(g_{t\phi}b^t + g_{\phi\phi}b^{\phi})}{u^r \rho} \\
      &+ \frac{b^r\frac{\partial u^r}{\partial r}(g_{t\phi}b^t + g_{\phi\phi}b^{\phi})}{(u^r)^2 \rho} -\frac{g_{\phi\phi}b^r\frac{\partial b^{\phi}}{\partial r}}{u^r \rho} - \frac{b^r b^t \frac{d g_{t\phi}}{dr}}{u^r \rho} -\frac{b^r b^{\phi}\frac{d g_{\phi\phi}}{dr}}{u^r \rho}-\frac{b^r(g_{t\phi}b^t +g_{\phi\phi}b^{\phi}\frac{d \Delta}{dr})}{2 \Delta u^r \rho},\\
      \mathcal{L}_v=&\frac{\partial h_{\rm tot}}{\partial v} u_{\phi}+ h_{\rm tot}\frac{\partial u_{\phi}}{\partial v}-\frac{g_{t\phi}b^r\frac{\partial b^t}{\partial v}}{u^r \rho}-\frac{g_{\phi\phi}b^r\frac{\partial b^{\phi}}{\partial v}}{u^r \rho}-\frac{\frac{\partial b^r}{\partial v}(g_{t\phi}b^t + g_{\phi\phi}b^{\phi})}{u^r \rho} + \frac{b^r\frac{\partial u^r}{\partial v}(g_{t\phi}b^t + g_{\phi\phi}b^{\phi})}{(u^r)^2 \rho}\\
      &-\frac{b^r(g_{t\phi}b^t + g_{\phi\phi}b^{\phi})}{u^r \rho v}- \frac{b^r v \gamma_{v}^2(g_{t\phi}b^t + g_{\phi\phi}b^{\phi})}{u^r \rho},\\
      \mathcal{L}_{\Theta}=& \frac{\partial h_{\rm tot}}{\partial \Theta}u_{\phi} -\frac{b^r(g_{t\phi}b^t + g_{\phi\phi}b^{\phi})}{2 \Theta u^r \rho},\\
      \mathcal{L}_{\lambda}=& \frac{\partial h_{\rm tot}}{\partial \lambda}u_{\phi} + h_{\rm tot}\frac{\partial u_{\phi}}{\partial \lambda} -\frac{g_{t \phi}b^r\frac{\partial b^t}{\partial \lambda}}{u^r \rho}-\frac{\frac{\partial b^r}{\partial \lambda}(g_{t\phi}b^t + g_{\phi\phi}b^{\phi})}{u^r \rho} - \frac{g_{\phi\phi}b^r\frac{\partial b^{\phi}}{\partial \lambda}}{u^r \rho}+\frac{\frac{\partial \mathcal{F}}{\partial \lambda}b^r(g_{t\phi}b^t + g_{\phi\phi}b^{\phi})}{2 \mathcal{F} u^r \rho}.      
\end{align*}

The radial momentum equation (Eq.~\ref{eq:radial momen1}) is written as,
\begin{equation*}
\mathcal{R}_r+\mathcal{R}_v \frac{d v}{dr}+\mathcal{R}_{\Theta}\frac{d \Theta}{dr}+\mathcal{R}_{\lambda}\frac{d \lambda}{dr}=0,
\end{equation*}
where the coefficients $\mathcal{R}_r$, $\mathcal{R}_v$, $\mathcal{R}_\theta$ and $\mathcal{R}_\lambda$ are given by,
\begin{align*}
    \mathcal{R}_r = & \frac{1}{h_{\rm tot} \left( \frac{1}{g_{rr}} + (u^r)^2 \right) \rho} \Biggl[-2b^r \frac{\partial b^r}{\partial r} - g_{rr} b^r \frac{\partial b^r}{\partial r} (u^r)^2   + \frac{1}{2} \left( \frac{1}{g_{rr}} + (u^r)^2 \right) + h_{\rm tot}u^r\frac{\partial u^r}{\partial r}\rho -\frac{3\left( \frac{1}{g_{rr}} + (u^r)^2 \right) \Theta \rho}{r \tau} \\
    & \frac{\frac{\partial \mathcal{F}}{\partial r}\left( \frac{1}{g_{rr}} + (u^r)^2 \right)\Theta \rho}{\tau \mathcal{F}}-b^ru^r\frac{\partial b^t}{\partial r}(g_{tt}u^t + g_{t\phi}u^{\phi})-b^ru^r\frac{\partial b^{\phi}}{\partial r }(g_{t\phi b^t} + g_{\phi\phi}b^{\phi})-\frac{(b^r)^2\frac{d g_{rr}}{dr}}{g_{rr}}-\frac{1}{2}(b^r)^2(u^r)^2\frac{d g_{rr}}{d r}\\
    & +\frac{h_{\rm tot}(u^r)^2 \rho \frac{d g_{rr}}{d r}}{2 g_{rr}}+ \frac{(b^t)^2\frac{d g_{tt}}{d r}}{2 g_{rr}} + \frac{b^t b^{\phi}\frac{ d g_{t \phi}}{d r}}{g_{rr}} -\frac{-(b^r)^2 \frac{d g_{\theta\theta}}{d r}}{2 g_{\theta \theta}} -(b^r)^2\left(\frac{g_{\phi\phi}\frac{d g_{tt}}{d r}}{2 (-g_{t\phi}^2 + g_{tt}g_{\phi\phi})} +\frac{g_{t\phi}\frac{d g_{t\phi}}{d r}}{2 (g_{t\phi}^2 - g_{tt}g_{\phi\phi})} \right)\\
    &-2  b^r u^r b^t (g_{tt}u^t + g_{t\phi}u^{\phi})\Biggr(\frac{g_{\phi\phi}\frac{d g_{tt}}{d r}}{2 (-g_{t\phi}^2 + g_{tt}g_{\phi\phi})} +\frac{g_{t\phi}\frac{d g_{t\phi}}{d r}}{2 (g_{t\phi}^2 - g_{tt}g_{\phi\phi})} \Biggr)-2  b^r u^r b^t (g_{t\phi}u^t + g_{\phi\phi}u^{\phi})\\
    &\Biggr(\frac{g_{t\phi}\frac{d g_{tt}}{d r}}{2 (g_{t\phi}^2 - g_{tt}g_{\phi\phi})}+\frac{g_{t t}\frac{d g_{t\phi}}{d r}}{2 (-g_{t\phi}^2 + g_{tt}g_{\phi\phi})} \Biggr) -g_{rr}(u^r)^2b^t\left(-\frac{b^t\frac{d g_{tt}}{d r}}{2 g_{rr}} -\frac{b^{\phi}\frac{d g_{t\phi}}{d r}}{2 g_{rr}}\right) + h_{\rm tot}u^t \rho\Biggr(-\frac{u^t\frac{d g_{tt}}{d r}}{2 g_{rr}} -\frac{u^{\phi}\frac{d g_{t\phi}}{d r}}{2 g_{rr}}\Biggr) \\
    &+\frac{(b^{\phi})^2\frac{d g_{\phi\phi}}{d r}}{ 2 g_{rr}} -2 b^r u^r b^{\phi}\left(g_{tt}u^t + g_{t \phi}u^{\phi}\right) \Biggr(\frac{g_{\phi\phi}\frac{d g_{t\phi}}{d r}}{2 (-g_{t\phi}^2 + g_{tt}g_{\phi\phi})}+\frac{g_{t \phi}\frac{d g_{\phi\phi}}{d r}}{2 (g_{t\phi}^2 - g_{tt}g_{\phi\phi})} \Biggr) -(b^r)^2\Biggr(\frac{g_{t\phi}\frac{d g_{t\phi}}{d r}}{2 (g_{t\phi}^2 - g_{tt}g_{\phi\phi})}\\
    &+\frac{g_{t t}\frac{d g_{\phi\phi}}{d r}}{2 (-g_{t\phi}^2 + g_{tt}g_{\phi\phi})} \Biggr)-2 b^r u^r b^{\phi}\left(g_{t\phi}u^t + g_{\phi \phi}u^{\phi}\right)\Biggr(\frac{g_{t\phi}\frac{d g_{t\phi}}{d r}}{2 (g_{t\phi}^2 - g_{tt}g_{\phi\phi})} +\frac{g_{t t}\frac{d g_{\phi\phi}}{d r}}{2 (-g_{t\phi}^2 + g_{tt}g_{\phi\phi})} \Biggr)\\
    & -g_{rr}(u^r)^2b^{\phi}\left(-\frac{b^t\frac{d g_{t\phi}}{d r}}{2 g_{rr}} -\frac{b^{\phi}\frac{d g_{\phi\phi}}{d r}}{2 g_{rr}}\right) + h_{\rm tot}u^t \rho\Biggr(-\frac{u^t\frac{d g_{t\phi}}{d r}}{2 g_{rr}} -\frac{u^{\phi}\frac{d g_{\phi\phi}}{d r}}{2 g_{rr}}\Biggr) - \frac{\left(\frac{1}{g_{rr}}+(u^r)^2\right)\Theta\rho\frac{d \Delta}{ dr}}{\tau \Delta}\Biggr],\\
    \mathcal{R}_v = & \frac{1}{h_{\rm tot} \left( \frac{1}{g_{rr}} + (u^r)^2 \right) \rho} \Biggl[ - b^r \frac{\partial b^r}{\partial r}-g_{rr}b^r \frac{\partial b^r}{\partial r} (u^r)^2 +\frac{1}{2}\frac{\partial (B^2)}{\partial v }\left(\frac{1}{g_{rr}}+(u^r)^2\right) + h_{\rm tot}u^r\frac{\partial u^r}{\partial v}\rho\\
    & -\frac{2\left(\frac{1}{g_{rr}}+(u^r)^2\right)\Theta \rho}{\tau v}-\frac{2\left(\frac{1}{g_{rr}}+(u^r)^2\right)\Theta \rho v \gamma_v^2}{\tau}-b^ru^r\frac{\partial b^t}{\partial v}(g_{tt}u^t+g_{t\phi}u^{\phi})-b^ru^r\frac{\partial b^{\phi}}{\partial v}(g_{t\phi}u^t+g_{\phi\phi}u^{\phi}) \Biggr], \\
    \mathcal{R}_{\Theta}=&\frac{1}{\tau h_{\rm tot}},\\
    %
    \mathcal{R}_{\lambda}=& \frac{1}{h_{\rm tot} \left( \frac{1}{g_{rr}} + (u^r)^2 \right) \rho}\Biggr[-2 b^r \frac{\partial b^r}{\partial \lambda} - g_{rr}b^r \frac{\partial b^r}{\partial \lambda}+\frac{1}{2}\left( \frac{1}{g_{rr}} + (u^r)^2 \right)\frac{\frac{\partial \mathcal{F}}{\partial \lambda}\left( \frac{1}{g_{rr}} + (u^r)^2 \right)\Theta\rho}{\tau \mathcal{F}}\\
    & -b^ru^r\frac{\partial b^{t}}{\partial \lambda}(g_{tt}u^t+g_{t\phi}u^{\phi})-b^ru^r\frac{\partial b^{\phi}}{\partial \lambda}(g_{t\phi}u^t+g_{\phi\phi}u^{\phi})\Biggr].
\end{align*}

\end{widetext}


\end{document}